# Laser Powder Bed Fusion of anisotropic Nd-Fe-B bonded magnets utilizing an in situ mechanical alignment approach

K. Schäfer*[a,f], R.G.T Fim*[a,b], F. Maccari[a,f], T. Braun[a,f], S. Riegg[a,f], K.Skokov[a,f], D. Koch[c], E. Bruder[d,f], I.Radulov[e,f] C. H. Ahrens[b], P. A. P. Wendhausen[b], O.Gutfleisch[a,f]

*Both authors contributed equally to this manuscript.

[a] *Functional Materials, Institute of Material Science, Technical University of Darmstadt, 64287 Darmstadt, Germany*

[b] *Mechanical Engineering Department, Federal University of Santa Catarina, Florianopolis, SC, 88040-001, Brazil*

[c] *Structural Research, Institute of Material Science, Technical University of Darmstadt, 64287 Darmstadt, Germany*

[d] *Physical Metallurgy, Institute of Materials Science, Technical University Darmstadt, 64287 Darmstadt, Germany*

[e] *Fraunhofer IWKS, Fraunhofer Research Institution for Materials Recycling and Resource Strategies, 63457 Hanau, Germany*

[f] *Additive Manufacturing Center, Technical University Darmstadt, 64289 Darmstadt, Germany*

*Corresponding authors: kilian.schaefer@tu-darmstadt.de and rafael.gitti@posgrad.ufsc.br

**Abstract**
Nd-Fe-B bonded magnets are an important class of permanent magnets, employed in many technological sectors. The Additive Manufacturing (AM) processes enables the fabrication of net-shape bonded magnets with complex geometries, allowing to tailor their magnetic stray field specifically for a given application. A crucial challenge to be addressed concerning AM of bonded magnets is the production of magnetically anisotropic components. The common approaches presented in the literature up to now, required a post-printing procedure or the complex integration of a magnetic field source into the AM process. Here, we present a technique to fabricate anisotropic bonded magnets via Laser Powder Bed Fusion (LPBF) by utilizing the mechanical alignment of anisotropic particles in a single step, without the need for a magnetic field source. Anisotropic bonded magnets were fabricated using a mixture of anisotropic Nd-Fe-B powder (MQA-38-14) and polyamide-12 (PA12). This magnetic powder consists of ellipsoidal particles, where the easy magnetization axis is distributed perpendicular to their longest side, which can be exploited to generate magnetic texture. Depending on the particle size used as feedstock, the degree of alignment ($<\cos\theta>$) can be tailored to a maximum of $<\cos\theta>$ = 0.78. The fabricated anisotropic bonded magnets exhibited a maximum remanence of $J_r$ = 377 mT and an energy product of $(BH)_{max}$ = 28.6 kJ/m³, respectively.

**Graphical Abstract:**

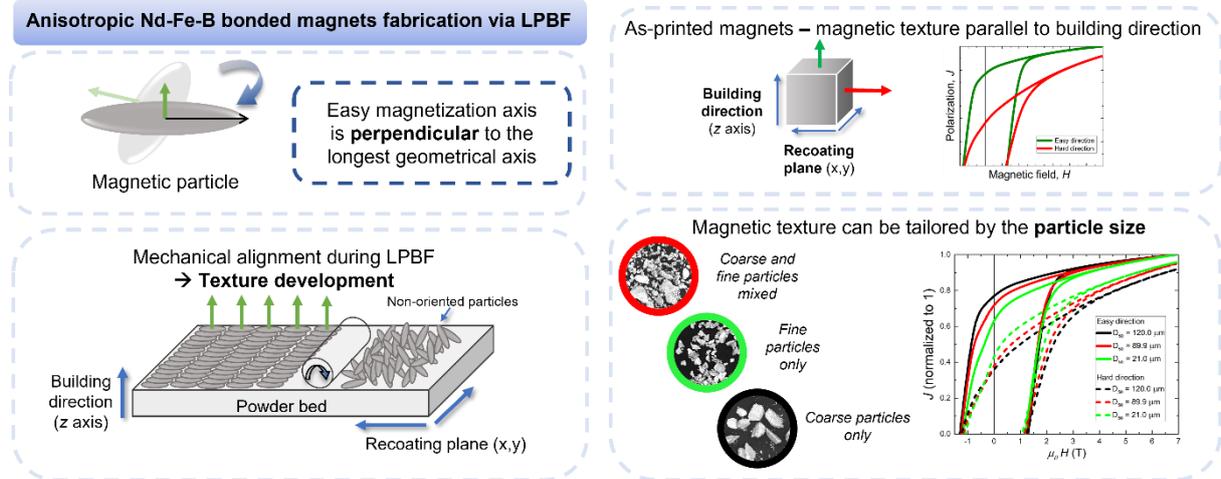

## 1. Introduction

Functional materials such as rare-earth (RE) based permanent magnets are essential for green energy conversion solutions (wind turbines and e-mobility) and high technology applications, including automation, sensors and actuators [1]. In recent years, Additive Manufacturing (AM) of permanent magnets has attracted great attention due to the opportunity produce functional components with tailored properties and complex geometries [2–6]. Conventionally, RE-based bonded magnets are manufactured by either compression or injection molding. Both techniques enable the fabrication of near-net-shape bonded magnets, but require specific tooling and molds for each design, limiting production capabilities of geometrical complexity and increasing costs [7]. In contrast, AM technologies enable the production of net-shape complex structures directly from a three-dimensional (3D) model. The layer-by-layer building principle eliminates the need for specific apparatus and further subtractive methods for geometrical adequation, in contrast to the conventional techniques. The flexibility of AM enables the production of complex-shaped bonded magnets with the possibility to tailor their magnetic stray field specifically for each application [8,9].

The main challenge for AM of bonded magnets is to increase the magnetic performance of the as-printed parts by the development of anisotropic magnets and porosity elimination. The figure of merit of a permanent magnet, such as remanent polarization ($J_r$) and maximum energy-product ($(BH)_{max}$) are proportional to the density and the degree texture of the magnet.

The two most investigated AM techniques for bonded magnets fabrication are extrusion-based AM and Powder-based AM. Up to now, the production of anisotropic bonded magnets is mainly reported for the first one, realized through two different approaches. The magnetic texture is developed as a post-printing operation or during the AM procedure (*in-situ*).

The first approach consists of heating up the as-printed bonded magnet close to the melting point of the polymeric binder in the presence of an external magnetic field, as first reported by Gandha et al. [7]. The viscosity reduction of the binder allows the magnetic particles to physically rotate and align in direction of the applied magnetic field. The advantage of this method is the possibility to make use of high external magnetic fields enabling the orientation of hard magnetic phases with large magnetocrystalline anisotropy, such as the $Nd_2Fe_{14}B$, $Sm_2Fe_{17}N_3$ and $SmCo_5$. Due to the texturization step being close to the melting point of the polymer, the used samples often exhibited geometric distortions [7].

The *in-situ* approach for texture development was first reported by Sonnleitner et al. for Fused Filament Fabrication (FFF) of strontium hexaferrite ($SrFe_{12}O_{19}$) and $Sm_2Fe_{17}N_3$ anisotropic bonded magnets [10]. In their work, the filaments with both magnetic fillers were extruded over the surface of a Nd-Fe-B permanent magnet, which provided an external magnetic field. Recently, Suppan et al. reported the production of anisotropic bonded magnets using a $Sm_2Co_{17}$ permanent magnet array attached to the FFF nozzle, allowing to tune the external magnetic field during printing [11]. The permanent magnet array provides an external magnetic field to align the magnetic filler during the process.

The fabrication of anisotropic bonded magnets via powder-based AM techniques, such as LPBF, are still unexplored in depth. So far, anisotropic bonded magnets production was realized only by the *in-situ* approach. Mapley et al. reported on the production of anisotropic Nd-Fe-B bonded magnets by integrating Helmholtz coils into the powder bed of an experimental setup [8]. The coil magnetizes the loose feedstock, made of 60 vol. % anisotropic MQA powder (Magnequench Int. – commercially available), promoting the alignment of the Nd-Fe-B particles. However, the applied field in the loose powder bed causes the displacement of the magnetic particles, generating voids on the powder bed that cannot be eliminated by the binder. This leads to highly porous samples (57% of porosity), reducing the beneficial contribution of the texture on the magnetic performance.

Recently, Schäfer et al. reported the preferential orientation of platelet-shaped Sm-Fe-N magnetic particles during the LPBF of bonded magnets [12]. During the recoating step, the Sm-Fe-N platelets are mechanically aligned parallel to the recoating direction due to the mechanical torque exerted by the recoater on the loose particles. This preferential alignment of the particles is related to the high aspect ratio of the platelets. Since the Sm-Fe-N platelets are magnetically isotropic, a magnetic texture could not be achieved. In the case of LPBF of fiber-reinforced composites, the preferential orientation of the fibers parallel to the recoating movement is known and it has a practical effect of mechanical anisotropy [13–16]. According to Chen et al., the fiber orientation depends on the mechanical torque exerted by the recoater over the fibers and can be tailored according to the recoating parameters [16].

To utilize the geometric alignment for the fabrication of anisotropic bonded magnets, a magnetic powder with a relation between morphology and magnetic anisotropy is needed. The commercial MQA powder meets these requirements. This powder is obtained via pulverization of hot deformed nanocrystalline magnets, which produces ellipsoidal particles with the easy magnetization axis perpendicular to the longest geometrical axis. The geometric alignment of each particle results in a magnetic texture in relation to the building direction of the LPBF process. Besides the absence of either post-printing procedures or in-situ magnetization arrays, this approach has the potential to avoid known phenomena such as geometrical distortions and high porosity. The use of a mechanical torque instead of a magnetic torque to physically align the magnetically anisotropic particles during the printing step (*in-situ*), is therefore an effective approach to obtain anisotropic bonded magnets.

## 2. Experimental
## 2.1. Feedstock characterization

The two powders used as feedstock for the LPBF process were MQA and PA12. Their properties are presented in Table 1. To evaluate the magnetic performance of the anisotropic powder, the MQA powder was embedded in epoxy, aligned in a system made of NdFeB magnets with a magnetic field of 1 T. Afterwards, isothermal magnetization curves were measured in the easy and hard direction with an external field of up to 7 T using a PPMS-VSM (Quantum Design GmbH). The alignment degree ($<\cos\theta>$) was calculated according to equation 1, based on the Fernengel function, adapted by Quispe, L.T. et al [17]:

$$< cos\theta > = \cos[\arctan(1.7263r)] \quad (1)$$

where $r$ is the remanent polarization ratio between the remanence in hard and easy magnetization directions ($r = J_r^{hard}/J_r^{easy}$). Magnetic domain structure analysis via Kerr microscopy (Zeiss Axio Imager.D2m evico magnetics GmbH) was performed on embedded and polished particles. To quantify the local crystallographic texture in a single MQA particle, electron backscatter diffraction (EBSD) was used in a FEG-SEM (Tescan Mira3), operating at 20 kV with a step size of 15 nm using neighbor pattern averaging and reindexing (EDAX NPAR[TM]) for post-processing.

For the fabrication of the bonded magnets with different particle sizes (section 3.1), the MQA powder was sieved into different particle sizes fractions. The sieving step was carried out using with a vibratory sieve shaker AS 200 Basic (RETSCH GmbH) in combination with a 63 µm test sieve (RETSCH GmbH). Each powder fraction was sieved twice. The particle size distributions (PSD) of the as received powder and the powder fractions above and below 63 µm were analyzed with a laser diffractometer Mastersizer 3000 (Malvern Instruments, Malvern). The Mastersizer 3000 uses laser light to obtain angular scattering intensity patterns. The angular scattering intensity data is then analysed to calculate the size of the particles that created the scattering pattern, using the Mie theory of light scattering. For the calculation of the PSD, the values for the refractive index and the absorption index of the characterised material were set to 1.3 and 0.5, respectively. SEM analysis (TESCAN Vega 3) was carried out to evaluate the morphology of the magnetic particles used in each feedstock. The flowability properties of the powder mixtures with the "as-received" and the sieved MQA were measured via the granular material flow analyzer GranuDrum (Granutools). With this device, the first avalanche angle was evaluated. The avalanche angle is the angle obtained by a linear regression of the surface at the maximum potential energy prior to the start of a powder avalanche [18]. For each powder the measurement was performed 10 times and the mean value and standard deviation are calculated. As a reference, the flowability of the pure PA12 was evaluated as well.

Table 1: Properties of the initial feedstock materials

| Powder | Particle size (µm) | Theoretical density (g/cm³) | Remanence $J_r$ (T) | Coercivity $\mu_0 H_c$ (T) |
|---|---|---|---|---|
| PA12 | 18 – 90 | 0.92 | - | - |
| MQA | 10 – 120 | 7.51 | 1.29 | 1.37 |

## 2.2. Fabrication of bonded magnets with LPBF

To prepare the feedstock for the LPBF process, the magnetic powder was mixed with the PA12 for 15 min using the shaking mixer TURBULA (WAB-GROUP). To optimize the density of the composites, the filler fraction of the MQA powder was varied between 40 to 65 volume % (vol. %) in 5 vol.% steps. For this series of experiments, the MQA powder was used in the "as-received" state. To investigate the dependence between the magnetic anisotropy induced during the process and the particle size, the sieved powders were used with a filling fraction of 55 vol. %. The LPBF process was performed on a Lisa Pro (Sinterit) device, equipped with a 5 W diode laser system (λ = 808 nm). A processing window analysis was carried out to determine the optimum laser parameters for porosity reduction. For the evaluation, the hatch spacing (HS) values were varied between 320 – 400 µm. Only the feedstocks composed of 65 and 40 vol. % of filler fractions were used for the assessment. After the determination of the processing window, the anisotropic bonded magnets were obtained using the processing parameters described in Tab.2.

Table 2: LPBF process parameters to fabricate the MQA - PA12 composites.

| Parameter | Powder surface temperature (°C) | Print chamber temperature (°C) | Laser scanning speed (m/s) | Laser spot size (mm) | Hatch distance (mm) | Layer height (mm) |
|---|---|---|---|---|---|---|
| Value | 177.5 | 140.0 | 0.054 | 0.4 | 0.32 - 0.4 | 0.125 |

## 2.3. Bonded magnet characterization

The sample geometry for analysis were cubes with a side length of 3 mm. The porosity level of the as-printed magnetic samples was determined by comparing the measured geometrical density with the calculated theoretical density for a fully dense sample. Estimations of the maximum achievable density of a pore-free as-printed bonded magnet (theoretical density of the composite, $\rho_{theo}$) can be calculated by the weighted sum of the volumetric fraction (f) of both components (magnet and polymer powder), expressed by equation 2:

$$\rho_{theo} = (f_{PA12} \times \rho_{PA12}) + (f_{MQA} \times \rho_{MQA}) \tag{2}$$

The filler fraction on the as-printed magnetic samples was analyzed via thermogravimetric measurements with a simultaneous thermal analyzer STA 409 CD (Netzsch, Germany) under nitrogen atmosphere. The temperature range of the measurement was 30 – 600 °C. The measurements were performed using samples produced with variable filler fractions and magnetic particle sizes.

To evaluate the alignment degree of the as-printed magnetic samples using both as-received and sieved MQA powder, isothermal magnetization curves at room temperature, with an external magnetic field up to 7 T, on both parallel and perpendicular to the building direction (z-axis) were carried out using a physical property measurement system coupled with vibrating sample magnetometer (PPMS-VSM - Quantum Design). The measurements were corrected using a demagnetization factor N = 0.2, which is also used in the datasheet of MQA from the manufacturer. The alignment degree ($<\cos\theta>$) was calculated according to equation 1.

XRD analysis was carried out in both surfaces (parallel and perpendicular to build direction) of the as-printed magnetic samples, using a Seifert PTS with Cu K$\alpha$ radiation and parallel beam optics to evaluate the resultant crystallographic orientation. SEM analysis (TESCAN Vega 3) of the as-printed bonded magnets in the same directions was performed to evaluate the geometric alignment of the MQA particles.

## 3. Results and discussion
### 3.1. Analysis of the initial properties – MQA-38-14

SEM images of the different MQA powders used as feedstock are presented in Figure 1 b), shows the MQA powder in the "as received" state whereas powder which was sieved below 63 µm can be seen in Figure 1 a) and the powder which was sieved above 63 µm is shown in Figure 1 c). Figure 2 shows the particle size distributions of the three powders. The "as-received" powder presents a $D_{50}$ = 81.1 mm (red line) and after the sieving process two powder fractions were obtained, with $D_{50}$ = 105.5 mm (black line, sieved above 63 mm) and with $D_{50}$ = 18.5 mm (green line, sieved below 63 mm), respectively.

Kerr microscopy reveals the correlation between particle morphology and crystallographic orientation. The easy magnetic direction is distributed perpendicular to the longest particle axis, as seen in Figure 3. The Kerr image in figure 3 a) of a few MQA particles shows the interaction domains which indicate the crystallographic orientation within the particles [19]. The domain pattern in particle I indicates a magnetic easy axis out of plane while the domain pattern in particle II indicates an in-plane easy axis. In both particles, the magnetic easy axis is perpendicular to the long axis of the particle. The domain pattern visible in different directions is schematically presented in figure 3 b).

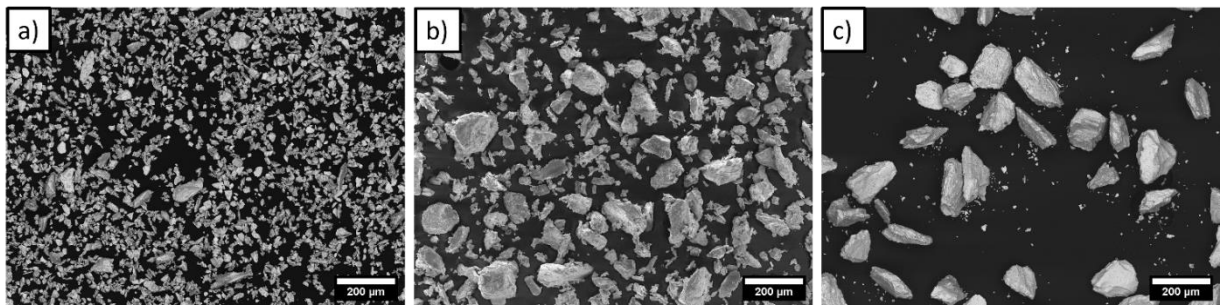

Figure 1: SEM images of the MQA powder with different particle sizes, where a) sieved powder below <63 µm ($D_{50}$ = 18.5 µm), (b) as-received condition ($D_{50}$ = 81.1 µm) and (c) sieved powder > 63 µm ($D_{50}$ = 105.5 µm)

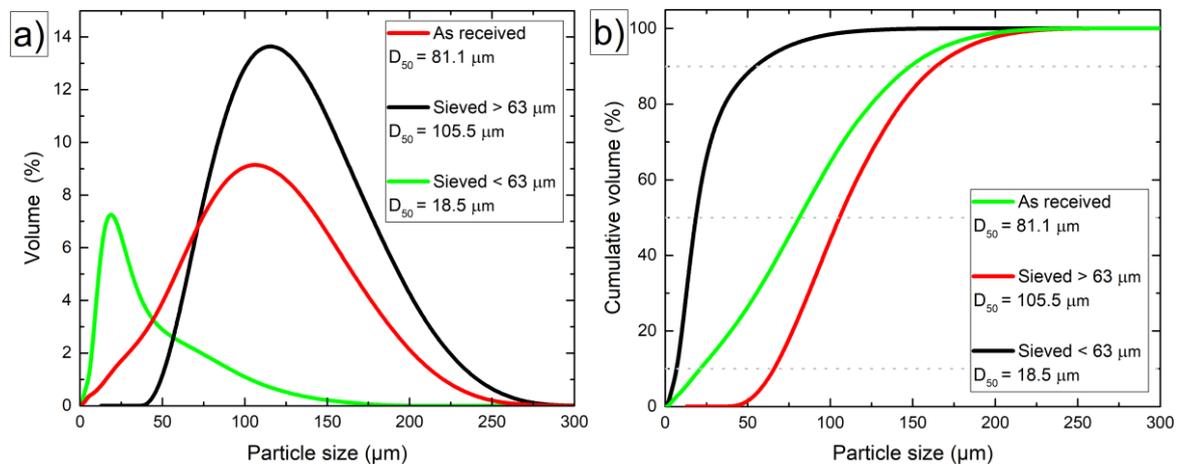

Figure 2: Particle size distributions of the as-received powder and the two sieved powders below and above 63 µm, where a) presents the volume fraction and b) presents the cumulative volume. The median of the cumulative distribution is the $D_{50}$ value.

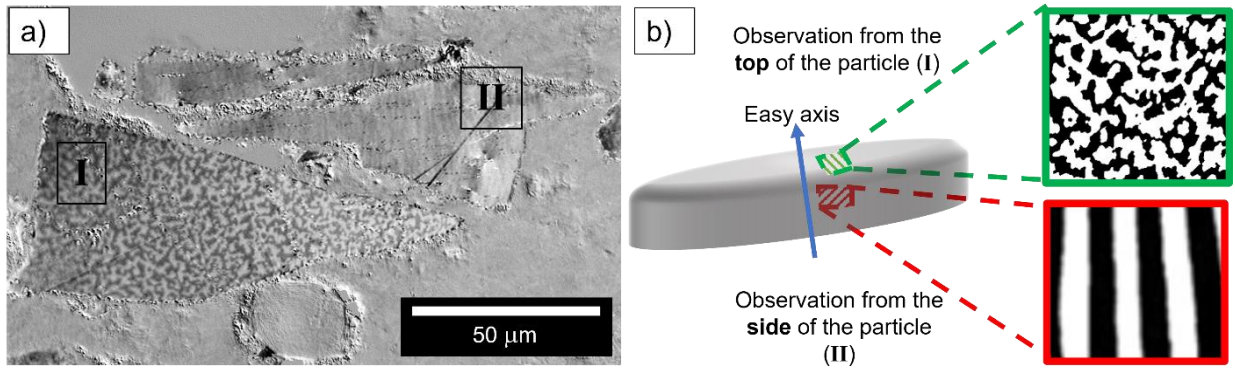

Figure 3: a) Kerr image of MQA powder particles. Interaction domains are visible which indicates the crystallographic orientation within the particles. The domain pattern in particle I indicates a magnetic easy axis out of plane while the domain pattern in particle II indicates an in-plane easy axis. In both particles, the magnetic easy axis is perpendicular to the long axis of the particle. The corresponding domain images of the direction out of plane and in plane are presented in b).The schematic domain structure representations are adopted from [19] and [20].

The powder characteristics can be attributed to the production method. The MQA is an anisotropic powder obtained from the pulverization of hot deformed magnets. The hot deformation process was first discovered by Lee et al. to produce anisotropic Nd-Fe-B magnets [21]. This method consists of the hot pressing of nanocrystalline Nd-Fe-B melt-spun ribbons into fully dense isotropic magnets followed by a hot deformation or die-upsetting step. This step causes the material to flow in the pressing direction leading to the crystallographic orientation of the $Nd_2Fe_{14}B$ grains, which is perpendicular to the deformation direction [22–25]. The anisotropic hot deformed magnets are then crushed into an anisotropic coarse powder, which can be used for anisotropic bonded magnets manufacturing. During the pulverization step, intergranular fractures occur preferentially along the flake boundaries [26], resulting in particles with ellipsoidal morphology.

Magnetic measurements at room temperature of the embedded and aligned magnetic powder (Section 2.1) are presented in Figure 4. Maximum remanence measured parallel to the easy magnetization direction is $J_r$ = 1.25 T, in accordance with the datasheet provided by the manufacturer ($J_r$ = 1.28 T). For the MQA powder, the <cosθ> value was calculated with equation 1 and the result is 0.95. A perfect alignment (<cosθ> = 1) is not achieved due to crystallographic misorientation of the grains mainly induced by the manufacturing process.

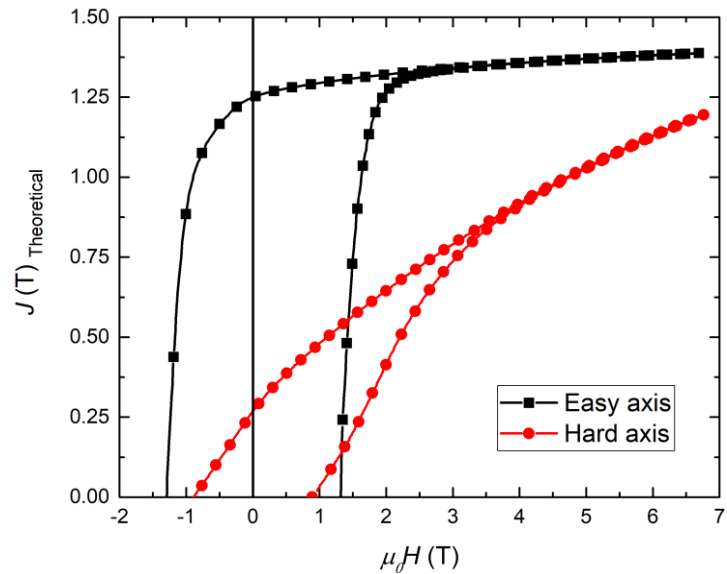

Figure 4: J(H) curves of easy and hard direction of the initial MQA feedstock, aligned before the measurement with a magnetic field of 1 T within epoxy. The polarization is calculated for a theoretical fully dense powder.

To study the grains and the crystallographic misalignment within the MQA powder particles, EBSD analysis in a single particle has been conducted. Figure 5 a) presents the inverse pole figure map of the investigated area using the center of the <001> direction as reference axis. Figure 5 b) shows the crystal direction map of the misalignment to the reference axis using the color coding shown in the misalignment histogram presented in Figure 5 c). The powder microstructure shows elongated grains with a thickness of around 50 nm and a length of 200 - 300 nm, which is typical for hot deformed $Nd_2Fe_{14}B$ [26]. According to the texture analysis, there is already misorientation present within the powder particles themselves.

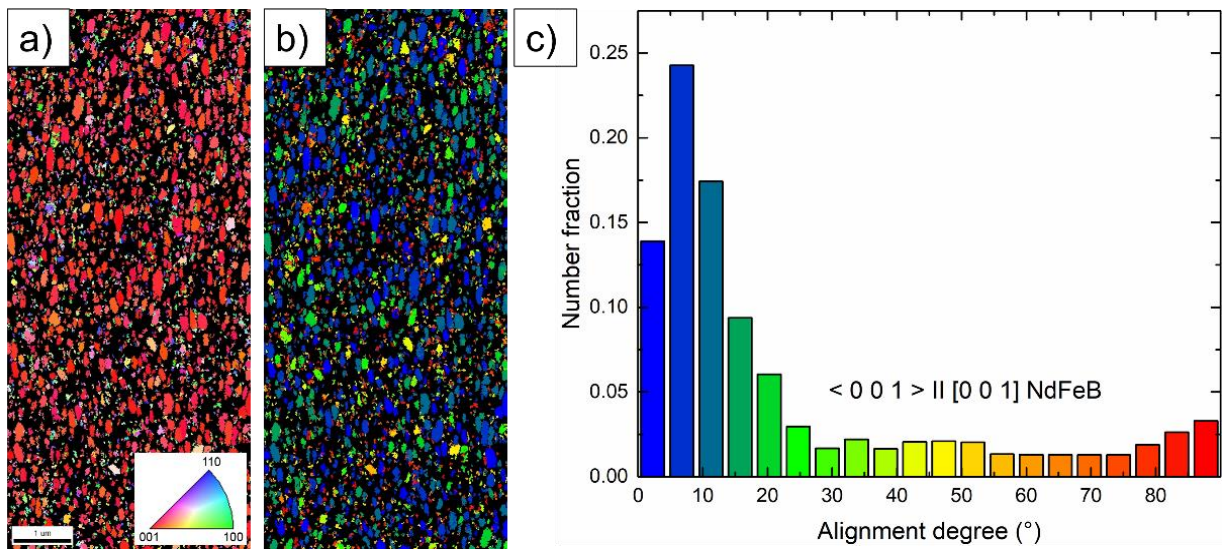

Figure 5: EBSD analysis of one MQA particle. a) Inverse pole figure map using the center of the <001> orientation distribution within the particle as reference axis. b) Crystal direction map showing the misalignment to the reference axis using the color coding provided in the misalignment histogram depicted in c). For interpretation of references to color in this figure, the reader is referred to the web version of this article.

For the LPBF process, the flowability of the feedstocks is highly important [27]. The flowability of a powder depends mainly on the particle size and morphology. The results of the first avalanche angle for the feedstocks consisting of 55 vol.% PA12 and the different MQA particle size fractions are presented in Figure 6 in comparison to the pure PA12. In a first approximation, it can be assumed that the larger the avalanche angle is, the poorer the flowability of the powder [18]. More detailed flowability data is in the appendix (Figure A 1 and Figure A 2). As visible in Figure 6, the first avalanche angle for the different feedstocks is the highest for the mixture with the smaller MQA particles and then decreases slightly for the mixture with as received powder. The avalanche angle of the mixtures is the lowest for the mixture with the larger MQA particles, but still larger than the value for the pure PA12, which serves here as a reference.

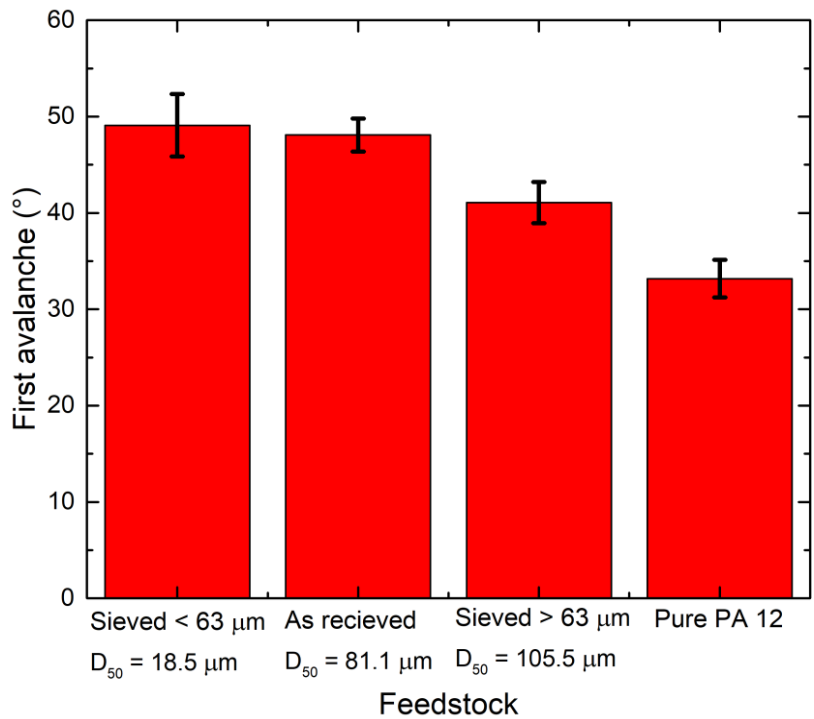

Figure 6: First avalanche angle for the feedstocks with 55 vol.% MQA with different particle size fractions in comparison to pure PA12 powder.

## 3.2. As-printed anisotropic magnetic samples

*3.2.1 Hatch spacing optimization*

As described in section 2.2, the hatch distance was optimized to obtain samples with a high density, which further increases the magnetic performance. Table 3 summarizes the mean geometrical ($\rho$), theoretical ($\rho_{theo}$) and relative density ($\rho_{rel}$), as well as the porosity values of the as-printed magnetic samples as a function of hatch spacing values (HS) for both 40 and 65 vol. % filler fraction. The theoretical density was calculated with Equation 2. The geometrical density $\rho$ and the relative densities of the samples values reached a maximum value when hatch spacing is HS = 400 μm. The ratio between measured density ($\rho$)/theoretical density ($\rho_{theo}$) represents the relative porosity level $\rho_{rel}$ of the as-printed samples, which is minimized for the HS = 400 μm process parameter. Therefore, the following experiments were performed with a HS value of 400 μm. Using HS = 320 μm, the as-printed magnetic samples presented geometrical distortions such as curling, mainly caused by the energy excess, and therefore they were not included in this analysis. Samples produced with the chosen HS values of 400 μm are presented in Figure 7, where a high shape accuracy is achieved.

Table 3: Geometrical density (r), relative density ($\rho_{theo}$) and porosity degree (100-$\rho_{rel}$) of the as-printed samples produced with 40 and 65 vol.% filler employing different hatch spacing (HS) values.

| Filler fraction (vol.%) | HS (μm) | Geometrical density, $\rho$ (g/cm³) | Theoretical density, $\rho_{theo}$ (g/cm³) | Relative density, $\rho_{rel}$ (%) | Mean porosity, 100-$\rho_{rel}$ (%) |
|---|---|---|---|---|---|
| 40 | 360 | 2.79 | 3.55 | 78 | 22 |
| 40 | 400 | 2.87 | 3.55 | 81 | 19 |
| 65 | 360 | 2.49 | 5.20 | 48 | 52 |
| 65 | 400 | 2.88 | 5.20 | 55 | 45 |

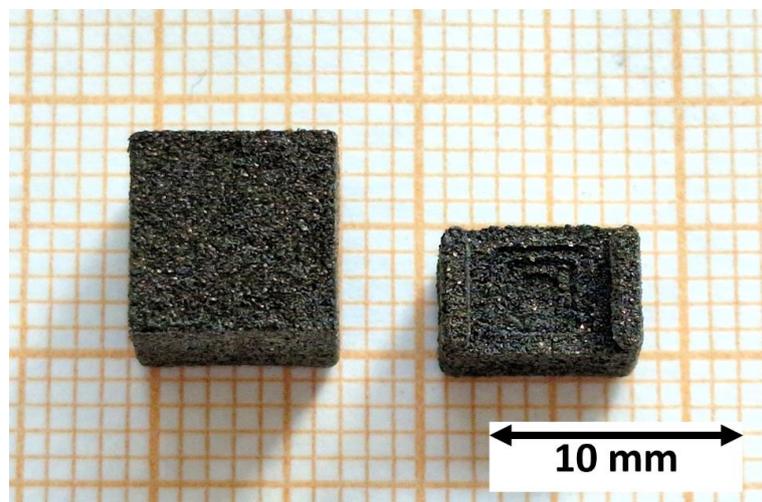

Figure 7: As-printed magnetic samples, a cube with 7.5 mm side length (left) and a more complex shape (right) which can be used within a wheel speed sensor [28].

*3.2.2 Influence of filler fraction on magnetic and microstructural properties*

Previous research demonstrated that the filler fraction is a crucial parameter for the resulting magnetic performance of bonded magnets [5,12]. As described in section 2.2, samples with filler fraction from 40 to 65 in steps of 5 vol.% were fabricated. To evaluate the actual filler fraction on the as-printed magnetic samples, TGA measurements were carried out and the results are shown in Figure 8 and summarized in Table 4. The PA12 degradation starts around 425°C and with an increasing temperature the binder is eliminated. The filler fraction on the as-printed magnetic samples is varying between 82 wt. % (40 vol.% MQA) and 93 wt. % (65 vol.% MQA), in good agreement with the nominal values as listed in Table **4**.

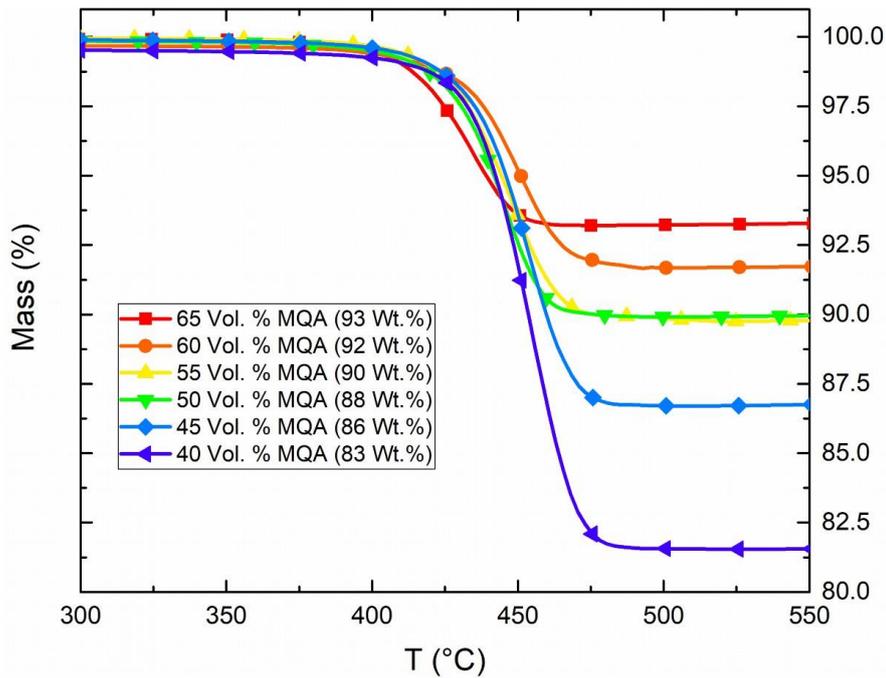

Figure 8: Mass as function of temperature during TGA measurement for the samples with different nominal MQA filler fraction.

Table 4: Comparison of nominal and measured filler fraction in vol.% and wt.%.

| **Nominal** | | **Measured** | |
| --- | --- | --- | --- |
| **vol.%** | **wt.%** | **vol.%** | **wt.%** |
| 65 | 93 | 65 | 93 |
| 60 | 92 | 60 | 92 |
| 55 | 90 | 55 | 90 |
| 50 | 88 | 55 | 90 |
| 45 | 86 | 47 | 87 |
| 40 | 83 | 38 | 82 |

Magnetic measurements at room temperature of the bonded magnets produced by LPBF was carried out. Figure 9 presents the J-H curves obtained from two directions, parallel and perpendicular to the building direction (z-axis). Remanent polarization ($J_r$) values are higher parallel to the z-axis of the as-printed magnets (building direction) than in the perpendicular direction (recoating plane), indicating a magnetic texture. $J_r$ values starts from $J_r^{easy}$ = 321 mT (65 vol. % filler), reaches a maximum of $J_r^{easy}$= 368 mT (50 vol. % filler) and then decreases to $J_r^{easy}$ = 326 mT (40 vol. % filler), respectively. Perpendicular to the building direction, the $J_r$ values for the same as-printed samples are $J_r^{hard}$ = 192 mT, 196 mT and 179 mT. The remanence values start to decrease in the range between 50 – 40 vol. % filler, mainly due to the high polymeric binder fraction [5]. The alignment degree values were determined with equation 1. The <cosθ> values range from 0.5 for an isotropic magnet to 1.0 for a perfectly oriented magnet. The as-printed magnetic samples exhibited similar alignment degree values, ranging from <cosθ> = 0.71 – 0.75, with a mean value of <cosθ> = 0.73.

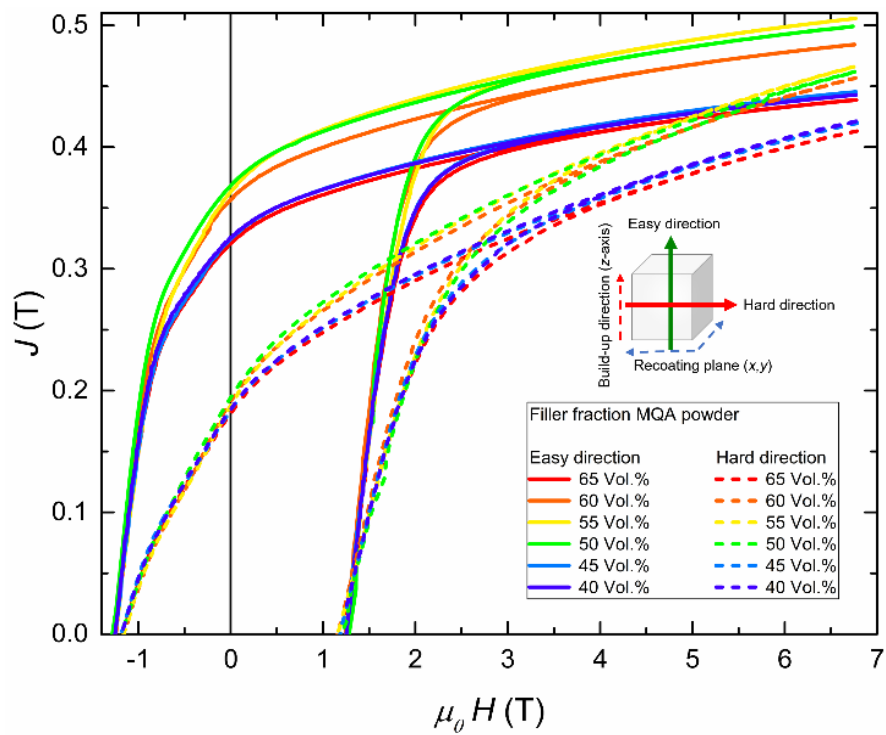

Figure 9: J(H) of the as printed samples as function of their MQA filler fraction for their easy (parallel to the building direction) and hard (perpendicular to building direction). For interpretation of references to color in this figure, the reader is referred to the web version of this article.

Figure 10 summarizes the geometric density, porosity and the alignment degree of the samples with different filling fractions. The geometrical (ρ), theoretical ($\rho_{theo}$) and relative density ($\rho_{rel}$), as well as the porosity are in appendix Table A 1, while the remanence values in easy and hard direction, the energy density and the alignment degree are in appendix Table A 2. Using the "as-received" MQA powder, the geometrical density increases first with increasing filler fraction, reaches maximum of ρ = 3.10 g/cm³ and then decreases with higher filler fractions. The porosity level (1-$\rho_{rel}$) increases with increasing filler fraction. Several factors can contribute to the porosity formation on the as-printed magnetic samples. The technological properties of the feedstock, such as the apparent density and flowability, which are related to the particle size, size distribution and morphology of the filler particles and processing parameters related to the sintering kinetics of the polymeric particles, such as the hatch spacing and laser scan speed play a role on porosity elimination [5,27]. Since the highest geometric density and degree of alignment is achieved with 55 vol.%, the next step was performed with this filler fraction.

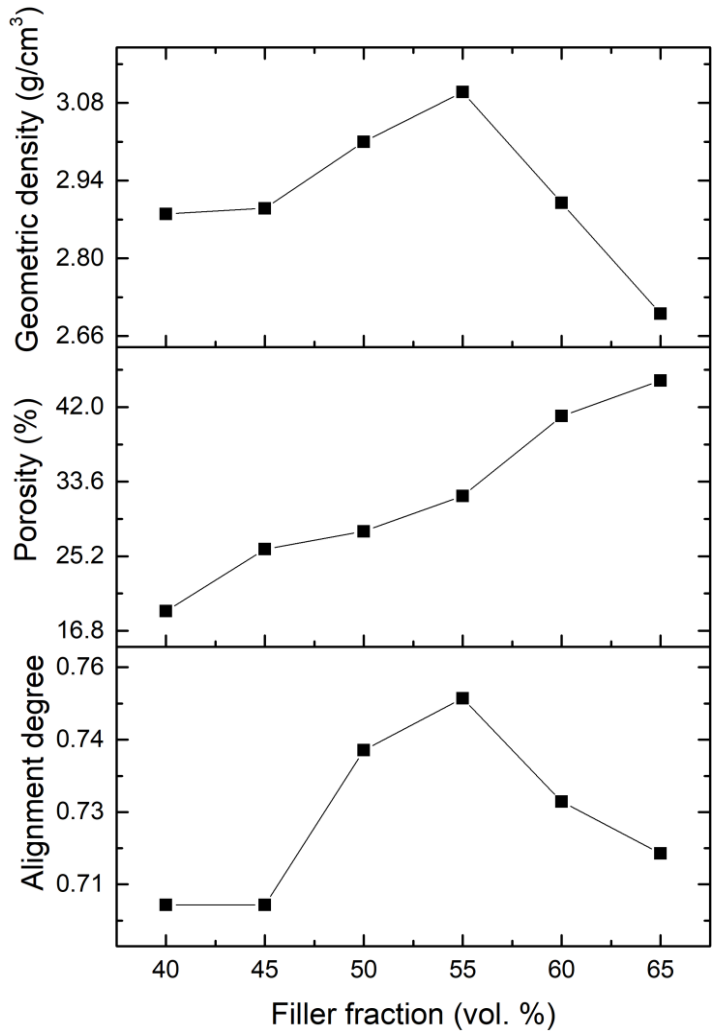

Figure 10: Geometric density, porosity and alignment degree of the samples with different filling fractions.

To understand the anisotropic magnetic behavior, the relation between morphology and geometric alignment of the MQA particles must be considered. As shown in the Kerr microscopy image in Figure 3, the easy magnetization axis of the $Nd_2Fe_{14}B$ phase is distributed perpendicular to the longest axis of the particles. During the recoating step, the MQA particles are mechanically oriented according to the recoating direction, with the longest axis horizontally aligned and the easy magnetization direction distributed in a plane that is perpendicular to the long geometrical axis. This particle orientation phenomenon was already observed by Schäfer et al. [12] for Sm-Fe-N bonded magnets obtained via LPBF. However, this effect did not create a magnetic texture since the used powder is magnetically isotropic. A mechanical alignment of non-spherical particles was also reported for fiber-reinforced polymeric composites obtained via LPBF [29].

SEM images from the as-printed magnetic samples using the as-received powder with a filler fraction of 55 vol.% is shown in Figure 11 where in (a) the area perpendicular to the building direction and in (b) the area parallel to the building direction, is shown respectively. In Figure 11 a), it is possible to observe the MQA particles distributed alongside the recoating plane (x,y), with the easy magnetization direction out of plane. A distinct microstructure is observed along the building direction, Figure 11 b). The geometric alignment induced during the printing step is visible in this image, where the MQA particles are distributed horizontally (parallel to recoating direction) and easy magnetization direction distributed perpendicular to the x,y-plane. A qualitative relation between the morphology and the magnetic axis could be also seen in the work of Schümann et al. [30].

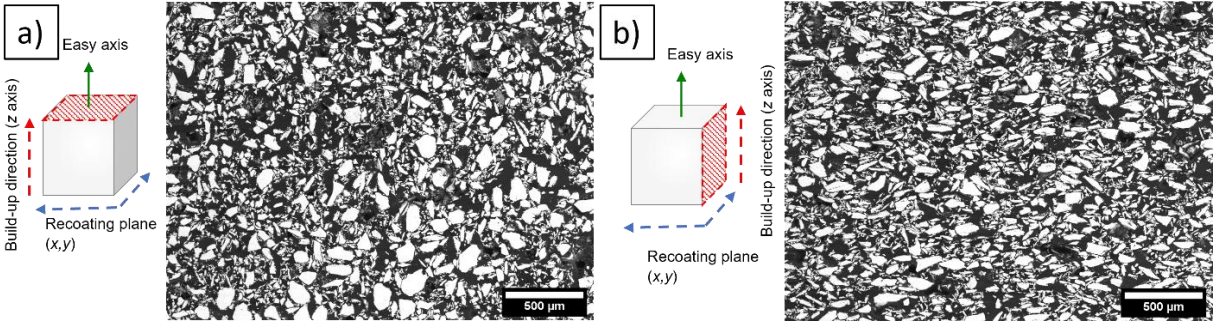

Figure 11: SEM-BSE image of the microstructure of the as-printed magnet. a) presents the view direction parallel to the building direction while b) shows the sample perpendicular to the building direction.

As additional method to investigate the texture, both surfaces analyzed via SEM (Figure **11**) were evaluated by XRD. Figure 12 presents the diffraction patterns from each surface, parallel (black lines) and perpendicular (red lines) to the building direction. For RE-Fe-B permanent magnets, the characteristic reflections for anisotropic magnets are the {00*l*} family. These planes are parallel to the easy magnetization direction (*c*-axis) and can be used for texture evaluation. In a magnetically anisotropic sample, the *c*-axis (easy magnetization direction) of a most of the $Nd_2Fe_{14}B$ grains is oriented in a particular direction, resulting in an increase on the intensity of the (00*l*) reflections, characterizing a crystallographic texture. As seen in Figure 12, the relative intensity from the (006) reflection parallel to the building direction (black line) is more intense than in the perpendicular direction (red line). This observation converges with both magnetic and microstructural analysis.

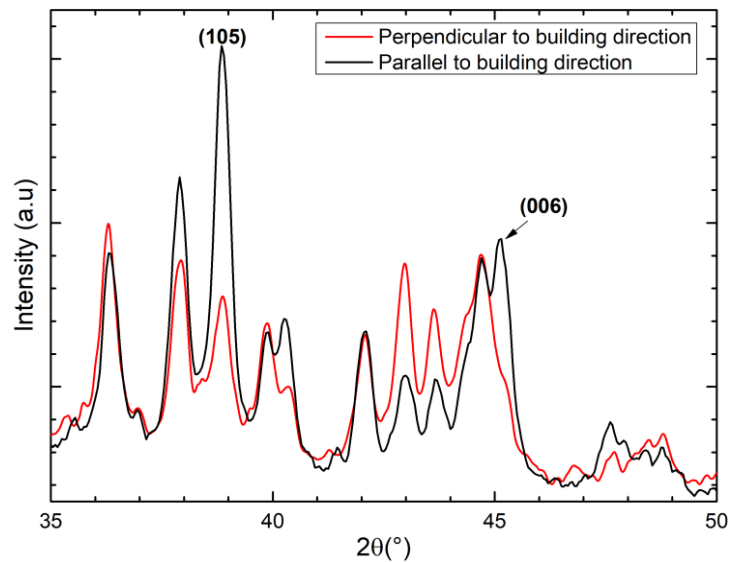

Figure 12: XRD patterns of the as-printed bonded magnet fabricated with 55% vol. MQA, from the perpendicular (red line) and parallel (black line) surfaces to the building direction.

### 3.3. Increasing the alignment degree: the particle size selection approach

Since the magnetic texture on the as-printed magnetic samples is induced by the physical orientation of the MQA particles during the LPBF step by mechanical torque, it could be possible to tailor the alignment degree of the printed samples by selecting particles with different morphologies and sizes. It was previously described in the literature, that the size and morphology are important parameters for the alignment of particles in composites [31]. For this purpose, two different types of anisotropic bonded magnets were produced with LPBF, using the optimum filler fraction (55 vol. %) described in section 3.1, using the MQA powder sieved in distinct particle sizes, as summarized in Table 2.

The porosity of the as-printed magnets was altered using different particle sizes, with a trend of increasing porosity with particle size reduction. The samples have a porosity of 33% ($D_{50}$ = 105.5 μm) and 50% ($D_{50}$ = 18.5 μm), respectively. The results are summarized in Table 5. The determined filler fractions by TGA, which are close to the nominal value are included in Table 5. The corresponding TGA graph is in the appendix (figure A 3). The effect of particle size on the porosity is clearly seen. Interparticle forces increases with decreasing particle size, reducing both the apparent density and flowability (Figure 6) of the powder, thus increasing the porosity level of the powder bed. Besides the porosity, the change of particle size also leads to variations on the alignment degree of the respective as-printed bonded magnets. Figure 13 a) presents the J(H) curves normalized to the highest polarization whereas the absolute values are shown in Figure 13 b). The results are summarized in Table 5. Remanent polarization ($J_r$) values measured parallel to the easy magnetization direction of the as-printed samples revealed an increase of $J_r^{easy}$ values when coarser particles are employed ($D_{50} = 105.5$ μm) in comparison to the as-received powder ($D_{50} = 81.1$ μm). The $J_r^{easy}$ value decreases when the particle size is reduced ($D_{50} = 18.5$ μm) mainly due to their higher porosity.

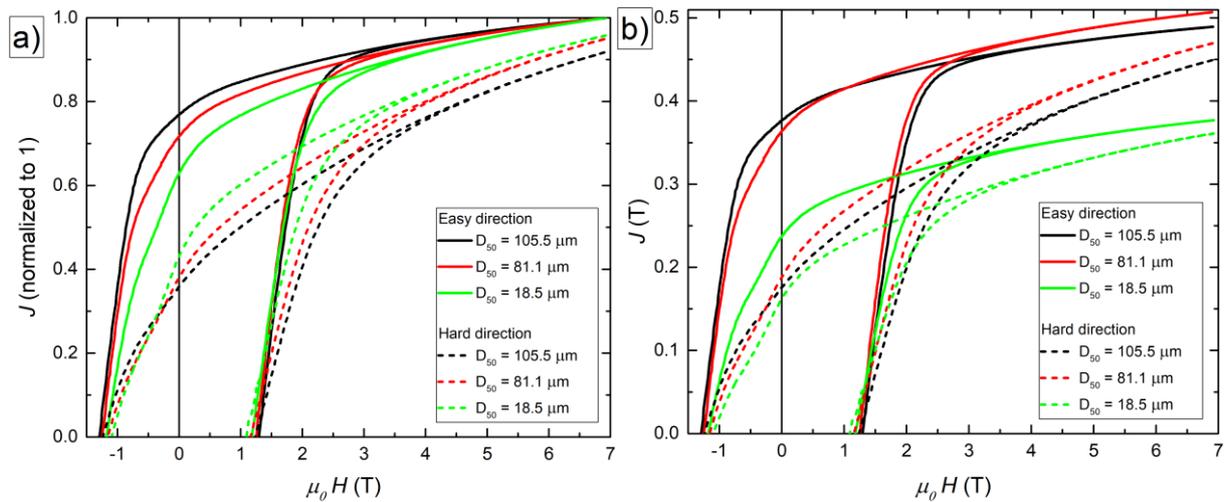

Figure 13: J(H) for the as printed samples with different MQA particle size distributions in easy and hard direction. a) the normalized values to the highest polarization and b) present the absolute values.

$J_r^{easy}$ values increases up to $J_r^{easy} = 377$ mT using the coarser powder with $D_{50} = 105.5$ μm and are reduced to $J_r^{easy} = 237$ mT using the finer powder with $D_{50} = 18.5$ μm. Alignment degree calculations revealed that $<\cos\theta>$ values reach a maximum of $<\cos\theta> = 0.78$ ($D_{50} = 105.5$ μm) for the coarser powder but are reduced to $<\cos\theta> = 0.64$ using the finer powder ($D_{50} = 18.5$ μm).

Table 5: Density and magnetic properties of the as-printed anisotropic bonded magnets with different magnetic particle size fractions as filler.

| Particle size $D_{50}$ (μm) | Filler fraction (vol.%) [wt.%] | Geometrical density ρ (g/cm³) | Porosity (%) | $J_r^{easy}$ (mT) | $J_r^{hard}$ (mT) | $(BH)_{max}$ (kJ/m³) | Alignment degree $<\cos\theta>$ |
|---|---|---|---|---|---|---|---|
| 105.5 | 53 [89] | 3.06 | 33 | 377 | 176 | 28.6 | 0.78 |
| 81.1 | 53 [89] | 3.10 | 34 | 363 | 188 | 26.8 | 0.75 |
| 18.5 | 57 [91] | 2.31 | 50 | 237 | 160 | 11.2 | 0.64 |

The results reveal the correlation between particle size and alignment degree induced during the spreading of a new powder layer. The magnetic powder in the "as-received" state presents a wide particle size distribution, as seen in Figure 2. The bonded magnets produced with this broad particle size distribution results in a $<\cos\theta>$ of 0.75. During the recoating step, the coarser fraction of the particles is aligned parallel to the recoater movement, inducing this magnetic anisotropy. However, the finer particles are less susceptible to mechanical alignment due to their tendency to agglomerate which is indicated by the decreased flowability properties of the small particles as visible in Figure 6. This reduces the overall <cosθ> to 0.65 values of the obtained as-printed samples with the finer particles. The fraction with the largest particles is susceptible to the mechanical alignment promoted during the recoating step, resulting in an increase on <cosθ> to 0.78. The geometrical alignment of the particles is visible via SEM analysis from cross sections of both directions (parallel and perpendicular to building directions) of the as-printed samples, using different powders, as seen on Figure 14, where (a,d) $D_{50}$ = 105.5 mm, (b,e) $D_{50}$ = 81.1 mm (as-received) and (c,f) $D_{50}$ = 18.5 mm, respectively.

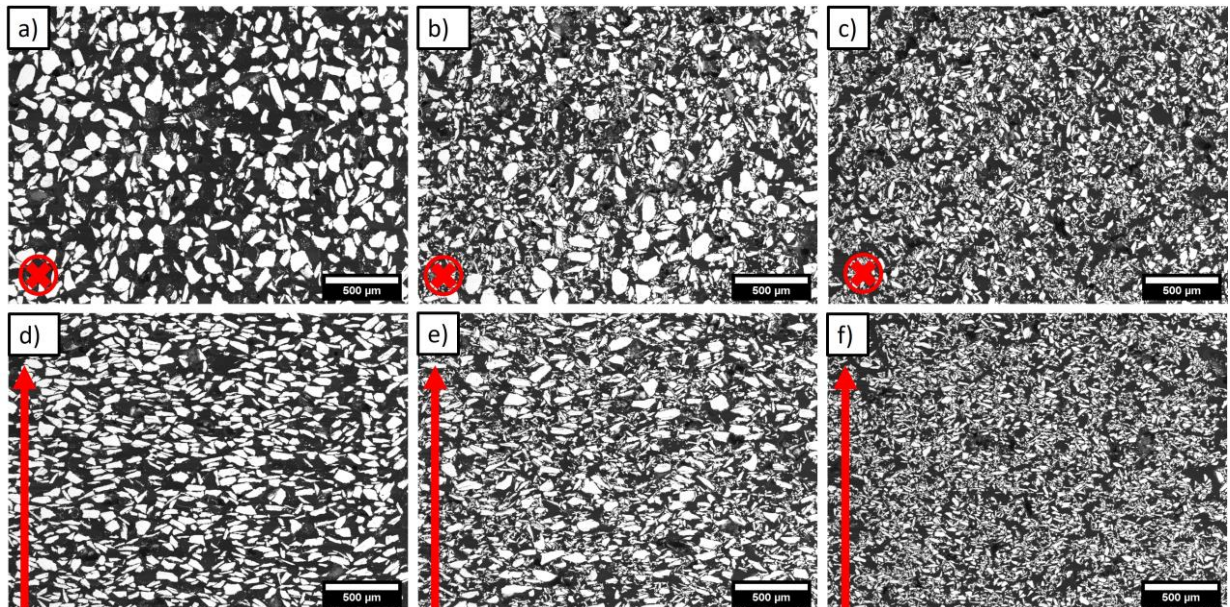

Figure 14: SEM-BSE images from the as-printed anisotropic bonded magnets, fabricated with different MQA particle sizes, where a,d) $D_{50}$ = 105.5 μm, b,e) $D_{50}$ = 81.1 μm and c,f) $D_{50}$ = 18.5 μm, respectively. The upper row presents images from the perpendicular to the building direction and the lower row from the parallel to the building direction. The red arrow is indicating the building direction.

From XRD patterns shown in Figure 15, it can be seen that the relative intensity of the (006) reflection varies according to the particle size used for bonded magnet manufacturing. Using the $D_{50}$ = 105.5 μm powder (black line) it is possible to observe an increase in the (006) relative intensity when compared to the as-printed magnet obtained using the $D_{50}$ = 81.1 μm powder (red line). This suggests that larger particles are more susceptible to the mechanical orientation during LPBF when compared to finer ones, directly affecting the texture development by this approach, in agreement with the magnetometry analysis.

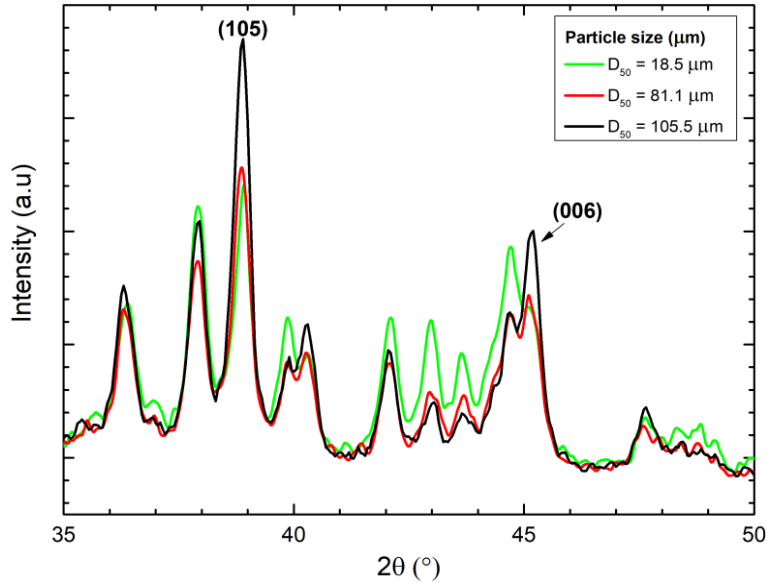

Figure 15: XRD patterns of the as-printed bonded magnets (55% vol. MQA) fabricated using different particle size distributions, where $D_{50}$ = 18.5 μm (green line), $D_{50}$ = 81.1 μm ("as-received powder", red line) and $D_{50}$ = 105.5 μm (green line), respectively. The XRD patterns are from the surface parallel to the building direction.

## 3.4 Magnetic versus mechanical torque assisted texture development – literature comparison

Figure 16 presents the remanence values of alignment degree (<cosθ>) and remanence of the as-printed anisotropic bonded magnets obtained in this work compared to several works from the literature, which employed the "*in-situ*" texture development as main approach.

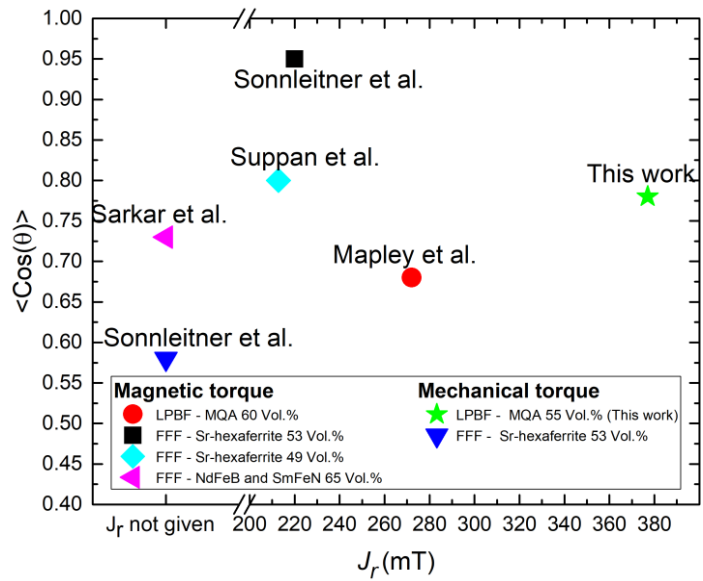

Figure 16: Alignment degree <cosθ> and remanence ($J_r$) values of this work in comparison with "in-situ" methods (magnetic or mechanical torque) from the literature. Remanence for mechanical alignment of Sonnleitner et al. [10] and Sarkar et al. [32] not given in literature. Other references: Mapley et al. [33], Suppan et al. [11].

As presented in the Introduction section, several approaches for "*in-situ*" magnetic texture development were evaluated. The use of a magnetic torque for particle orientation, however, is limited when considering some key aspects. The first one is the need for a magnetic field source integration into the experimental setup, providing a homogeneous stray field in a determined volume. This source can be a sintered permanent magnet used as substrate or the integration of a permanent magnet array to the AM equipment. Using a permanent magnet as printing substrate has the disadvantage of producing a homogeneous stray field in a very small region close to the surface, limiting the dimensions of the magnet to be manufactured [10]. The integration of a magnetic field source into the FFF equipment, for example, led to geometrical distortions on the as-printed bonded magnets [11]. Depending on the magnitude of the external magnetic field, the molten filament can be displaced during printing, inducing such defect.

In the case of Powder-based AM, the magnetization of the loose powder bed for texture development is also challenging, as reported by Mapley et al. [33]. The displacement of the anisotropic particles when subject to external magnetic fields can limit the density of the as-printed bonded magnets, reducing the effectiveness of magnetic texture on the overall magnetic properties.

The physical orientation of the anisotropic magnetic particles using a mechanical torque instead a magnetic, presented in this work, shows some advantages over the cited above. The first one is the absence of integration or use of a magnetic field to promote this alignment. The particle size seems to cause a more sensible variation on the alignment degree values than other explored variables, such as the printing parameters (HS) and feedstock composition. The contribution of these parameters on $<\cos\theta>$ needs a deeper investigation.

The improvement of $<\cos\theta>$ values was demonstrated using only the largest particles while maintaining other important characteristics, such as porosity level and geometrical features, in comparison to the samples produced with the as received powder.

In terms of absolute values of alignment degree, as-printed $SrFe_{12}O_{19}$ bonded magnets fabricated via FFF exhibited values ranging from 0.75 [11] to 0.96 [10]. In the case of Nd-Fe-B and Nd-Fe-B/Sm-Fe-N hybrid bonded magnets, the as-printed components presented values varying between 0.75 [10] and 0.83 [32], respectively.

In the case of LPBF, the alignment degree values of the as-printed Nd-Fe-B bonded magnets were limited to the range of $<\cos\theta> = 0.57 – 0.68$ [33]. The physical alignment of the anisotropic particles, reported in this work, allowed the fabrication of Nd-Fe-B bonded magnets with a mean alignment degree of $<\cos\theta> = 0.73$, further increased up to $<\cos\theta> = 0.78$. This represents a 14% increase on $<\cos\theta>$ compared to the reported values in the literature [33], employing the same AM technique and raw materials. The bonded magnets fabricated with the largest particles have a remanence of 377 mT, which exceeds comparable literature values in the field of *in situ* alignment methods by 39 %.

## 4. Summary and outlook

In this work, the fabrication of anisotropic bonded magnets with LPBF by mechanical alignment was investigated. It was shown, that if a magnetic powder with a relationship between particle morphology and magnetic anisotropy is used, the fabrication of anisotropic bonded magnets is possible. The commercially available magnetic powder MQA has this property since the ellipsoid shaped particles, indicated by Kerr microscopy, show the orientation of the magnetic easy axis is perpendicular to the largest dimension of the flake. The MQA filler fraction, in the investigated range of 40 – 65 vol.%, has only a minor influence on the degree of alignment. On the other hand, the particle size has a clear influence since finer particles have the tendency to agglomerate, which results in a decreased level of mechanical alignment. The degree of alignment can be increased to $<\cos \theta> = 0.78$ if only the largest particle size fraction with $D_{50} = 105.5$ µm is selected. The obtained magnets reveal a remanence of 377 mT, which exceeds comparable literature values in the field of *in situ* alignment methods by 39 %.

Our results illustrate the potential of producing anisotropic bonded magnets with simultaneously high geometrical accuracy and magnetic performance. In future works, this mechanism can be exploited for local tailoring of the stray field. It is possible to produce bonded magnets with layers with different particle sizes and therefore different degrees of alignment which results in different magnetic stray field strengths. This could be used for stray field designs specifically for an application. Also, larger magnetic particles could be used to enhance the effect of mechanical alignment. However, here optimization is necessary between a high degree of alignment and decreasing processability. In addition, the here presented alignment approach should be combined with an in-situ magnetic field to achieve larger degrees of alignment.

## Acknowledgments

This work was financially supported by the Deutsche Forschungsgemeinschaft (DFG, German Research Foundation), Project ID No. 405553726, TRR 270, the Deutscher Akademischer Austauschdienst (DAAD) and Brazilian CAPES for the Co-Financed Short Term Research Grant 2021 No. 57552433 (Fim, R.G.T).

**Appendix:**

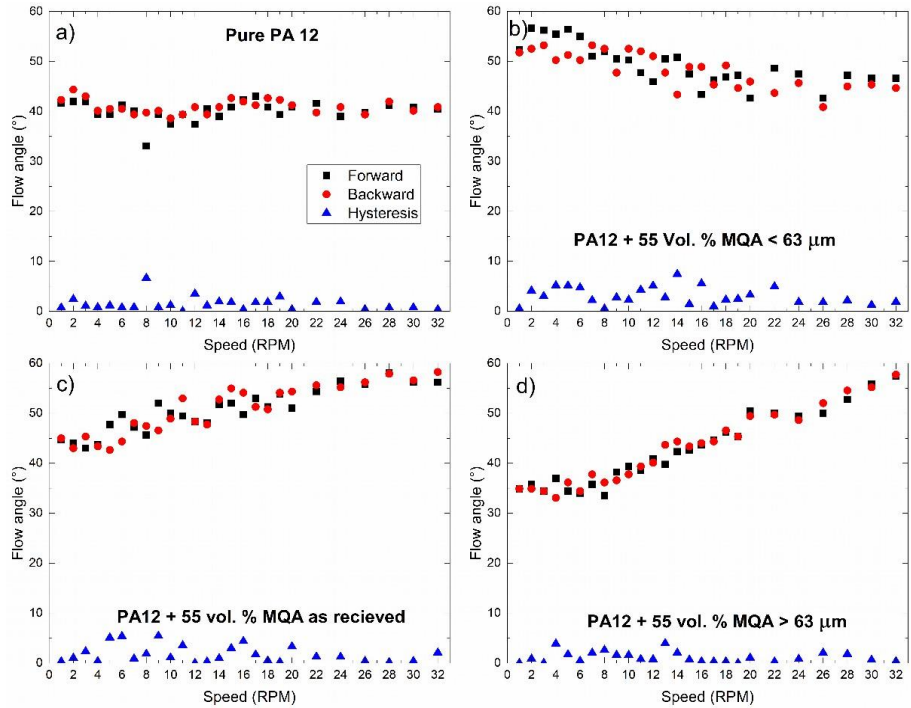

Figure A 1: Flow angle of a) pure PA12 and the feedstocks with different particle size distributions: b) Sieved MQA < 63 μm with $D_{50}$ = 21.0 μm, c) As received with $D_{50}$ = 89.6 μm and d) Sieved MQA > 63 μm with $D_{50}$ = 120.0 μm.

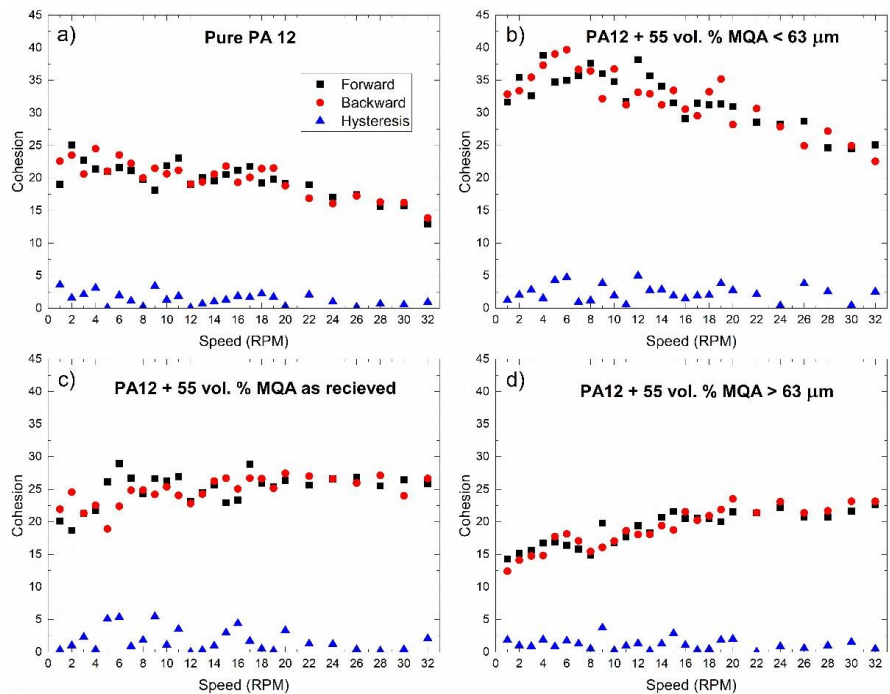

Figure A 2: Cohesion of a) pure PA12 and the feedstocks with different particle size distributions: b) Sieved MQA < 63 μm with $D_{50}$ = 21.0 μm, c) As received with $D_{50}$ = 89.6 μm and d) Sieved MQA > 63 μm with $D_{50}$ = 120.0 μm.

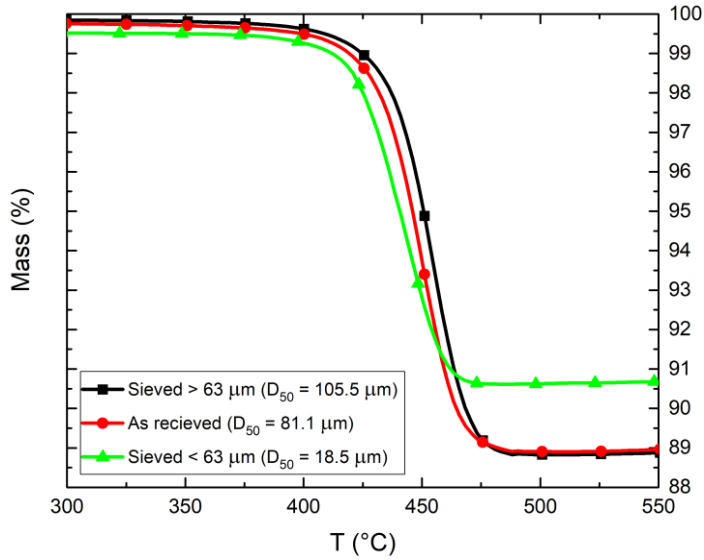

Figure A 3: Mass as function of the temperature from thermogravimetric analysis of the bonded magnets fabricated with different MQA particle size fractions with nominal 55 vol.% (90 wt.%).

Table A 1: Geometrical density ($\rho$), theoretical density ($\rho_{theo}$), relative density ($\rho_{rel}$) and mean porosity (100-$\rho_{rel}$) values of the as-printed bonded magnets fabricated with different filler fractions.

| Nominal filler fraction (vol.%) | Geometrical density, $\rho$ (g/cm³) | Theoretical density, $\rho_{theo}$ (g/cm³) | Relative density, $\rho_{rel}$ (%) | Porosity, 100-$\rho_{rel}$ (%) |
|---|---|---|---|---|
| 65 | 2.88 | 5.20 | 55 | 45 |
| 60 | 2.90 | 4.87 | 59 | 41 |
| 55 | 3.10 | 4.54 | 68 | 32 |
| 50 | 3.01 | 4.21 | 72 | 28 |
| 45 | 2.89 | 3.88 | 74 | 26 |
| 40 | 2.87 | 3.55 | 81 | 19 |

Table A 2: Remanence values of the as-printed magnets measured in easy ($J_r^{easy}$) and hard direction ($J_r^{hard}$) and the corresponding calculated alignment degree.

| Filler fraction (% vol.) | $J_r^{easy}$ (mT) | $J_r^{hard}$ (mT) | (BH)$_{max}$ (kJ/m³) | Alignment degree, <cos θ> |
|---|---|---|---|---|
| 65 | 321 | 181 | 20.5 | 0.72 |
| 60 | 357 | 191 | 25.3 | 0.73 |
| 55 | 363 | 183 | 26.8 | 0.75 |
| 50 | 368 | 194 | 27.4 | 0.74 |
| 45 | 324 | 184 | 20.9 | 0.71 |
| 40 | 326 | 185 | 21.2 | 0.71 |

References

[1] O. Gutfleisch, M.A. Willard, E. Brück, C.H. Chen, S.G. Sankar, J.P. Liu, Magnetic materials and devices for the 21st century: stronger, lighter, and more energy efficient, Adv. Mater. Weinheim. 23 (2011) 821–842. https://doi.org/10.1002/adma.201002180.

[2] C. Huber, 3D Printed Polymer-Bonded NdFeB Magnets for a Tailored Magnetic Field. PhD Thesis, Vienna (2017)

[3] Kinjal Gandha, Ikenna C. Nlebedim, Vlastimil Kunc, Edgar Lara-Curzio, Robert Fredette, M. Parans Paranthaman, Additive manufacturing of highly dense anisotropic Nd-Fe-B bonded magnets.

[4] K. Gandha, M.P. Paranthaman, B.C. Sales, H. Wang, A. Dalagan, T.N. Lamichhane, D.S. Parker, I.C. Nlebedim, 3D printing of anisotropic Sm–Fe–N nylon bonded permanent magnets, Engineering Reports (2021). https://doi.org/10.1002/eng2.12478.

[5] R. Fim, A.A. Mascheroni, L.F. Antunes, J. Engerroff, C.H. Ahrens, P. Wendhausen, Increasing packing density of Additively Manufactured Nd-Fe-B bonded magnets, Additive Manufacturing 35 (2020) 101353. https://doi.org/10.1016/j.addma.2020.101353.

[6] M. Röhrig, R.G.T. Fim, R.N. de Faria, C.C. Plá Cid, C.H. Ahrens, P. Wendhausen, 2022. Laser Powder Bed Fusion of Sm-Fe-N Bonded Magnets Employing Flake Powders. 3D Printing and Additive Manufacturing, 3dp.2021.0228. https://doi.org/10.1089/3dp.2021.0228.

[7] K. Gandha, L. Li, I.C. Nlebedim, B.K. Post, V. Kunc, B.C. Sales, J. Bell, M.P. Paranthaman, Additive manufacturing of anisotropic hybrid NdFeB-SmFeN nylon composite bonded magnets, Journal of Magnetism and Magnetic Materials 467 (2018) 8–13. https://doi.org/10.1016/j.jmmm.2018.07.021.

[8] C. Huber, M. Goertler, C. Abert, F. Bruckner, M. Groenefeld, I. Teliban, D. Suess, Additive Manufactured and Topology Optimized Passive Shimming Elements for Permanent Magnetic Systems, Sci. Rep. 8 (2018) 14651. https://doi.org/10.1038/s41598-018-33059-w.

[9] W. Kersten, L. Brandl, R. Wagner, C. Huber, F. Bruckner, Y. Hasegawa, D. Suess, S. Sponar, Additive-Manufactured and Topology-Optimized Permanent-Magnet Spin Rotator for Neutron Interferometry, Phys. Rev. Applied 12 (2019). https://doi.org/10.1103/PhysRevApplied.12.014023.

[10] K. Sonnleitner, C. Huber, I. Teliban, S. Kobe, B. Saje, D. Kagerbauer, M. Reissner, C. Lengauer, M. Groenefeld, D. Suess, 3D printing of polymer-bonded anisotropic magnets

in an external magnetic field and by a modified production process, Appl. Phys. Lett. 116 (2020) 92403. https://doi.org/10.1063/1.5142692.

[11] M. Suppan, C. Huber, K. Mathauer, C. Abert, F. Brucker, J. Gonzalez-Gutierrez, S. Schuschnigg, M. Groenefeld, I. Teliban, S. Kobe, B. Saje, D. Suess, In-situ alignment of 3D printed anisotropic hard magnets, Sci. Rep. 12 (2022) 17590. https://doi.org/10.1038/s41598-022-20669-8.

[12] K. Schäfer, T. Braun, S. Riegg, J. Musekamp, O. Gutfleisch, Polymer-bonded magnets produced by laser powder bed fusion: Influence of powder morphology, filler fraction and energy input on the magnetic and mechanical properties, Materials Research Bulletin 158 (2023) 112051. https://doi.org/10.1016/j.materresbull.2022.112051.

[13] A. Jansson, L. Pejryd, Characterisation of carbon fibre-reinforced polyamide manufactured by selective laser sintering, Additive Manufacturing 9 (2016) 7–13. https://doi.org/10.1016/j.addma.2015.12.003.

[14] R.B. Floersheim, G. Hou, K. Firestone, CFPC material characteristics and SLS prototyping process, RPJ 15 (2009) 339–345. https://doi.org/10.1108/13552540910993860.

[15] S. Arai, S. Tsunoda, A. Yamaguchi, T. Ougizawa, Effects of short-glass-fiber content on material and part properties of poly(butylene terephthalate) processed by selective laser sintering, Additive Manufacturing 21 (2018) 683–693. https://doi.org/10.1016/j.addma.2018.04.019.

[16] C. Badini, E. Padovano, R. de Camillis, V.G. Lambertini, M. Pietroluongo, Preferred orientation of chopped fibers in polymer-based composites processed by selective laser sintering and fused deposition modeling: Effects on mechanical properties, J. Appl. Polym. Sci. 137 (2020) 49152. https://doi.org/10.1002/app.49152.

[17] L.T. Quispe, M. Röhrig, J.L. Monteiro, L.U. Lopes, P.A. Wendhausen, Fast experimental estimation of texture distribution and alignment degree in permanent magnets, Journal of Applied Physics 128 (2020) 165101. https://doi.org/10.1063/5.0028005.

[18] A. Amado, M. Schmid, G. Levy, K. Wegener, Advances in SLS Powder Characterization, University of Texas at Austin, 2011. http://dx.doi.org/10.26153/tsw/15306

[19] K. Khlopkov, O. Gutfleisch, D. Hinz, K.-H. Müller, L. Schultz, Evolution of interaction domains in textured fine-grained Nd2Fe14B magnets, Journal of Applied Physics 102 (2007) 23912. https://doi.org/10.1063/1.2751092.

[20] D. Goll, R. Loeffler, J. Herbst, R. Karimi, U. Pflanz, R. Stein, G. Schneider, Novel Permanent Magnets by High-Throughput Experiments, JOM 67 (2015) 1336–1343. https://doi.org/10.1007/s11837-015-1422-8.

[21] R. Lee, E. Brewer, N. Schaffel, Processing of Neodymium-Iron-Boron melt-spun ribbons to fully dense magnets, IEEE Trans. Magn. 21 (1985) 1958–1963. https://doi.org/10.1109/TMAG.1985.1064031.

[22] A. Kirchner, D. Hinz, V. Panchanathan, O. Gutfleisch, K.H. Muller, L. Schultz, Improved hot workability and magnetic properties in NdFeCoGaB hot deformed magnets, IEEE Trans. Magn. 36 (2000) 3288–3290. https://doi.org/10.1109/20.908772.

[23] A. Kirchner, J. Thomas, O. Gutfleisch, D. Hinz, K.-H. Müller, L. Schultz, HRTEM studies of grain boundaries in die-upset Nd–Fe–Co–Ga–B magnets, Journal of Alloys and Compounds 365 (2004) 286–290. https://doi.org/10.1016/S0925-8388(03)00661-3.

[24] T. Nishio, Y. Kasai, V. Panchanathan, J.J. Croat, Effect of Co content on magnetic properties of hot-worked Nd-Dy-Ce-Fe-Co-B magnets, IEEE Trans. Magn. 28 (1992) 2853–2855. https://doi.org/10.1109/20.179649.

[25] T. Saito, M. Fujita, T. Kuji, K. Fukuoka, Y. Syono, The development of high performance Nd–Fe–Co–Ga–B die upset magnets, Journal of Applied Physics 83 (1998) 6390–6392. https://doi.org/10.1063/1.367522.

[26] John J. Croat, Modern Permanent magnets (2022).

[27] A.B. Spierings, M. Voegtlin, T. Bauer, K. Wegener, Powder flowability characterisation methodology for powder-bed-based metal additive manufacturing, Prog Addit Manuf 1 (2016) 9–20. https://doi.org/10.1007/s40964-015-0001-4.

[28] C. Huber, C. Abert, F. Bruckner, M. Groenefeld, O. Muthsam, S. Schuschnigg, K. Sirak, R. Thanhoffer, I. Teliban, C. Vogler, R. Windl, D. Suess, 3D print of polymer bonded rare-earth magnets, and 3D magnetic field scanning with an end-user 3D printer, Appl. Phys. Lett. 109 (2016) 162401. https://doi.org/10.1063/1.4964856.

[29] H. Chen, W. Zhu, H. Tang, W. Yan, Oriented structure of short fiber reinforced polymer composites processed by selective laser sintering: The role of powder-spreading process, International Journal of Machine Tools and Manufacture 163 (2021) 103703. https://doi.org/10.1016/j.ijmachtools.2021.103703.

[30] M. Schümann, D.Y. Borin, S. Huang, G.K. Auernhammer, R. Müller, S. Odenbach, A characterisation of the magnetically induced movement of NdFeB-particles in magnetorheological elastomers, Smart Mater. Struct. 26 (2017) 95018.

[31] R.M. Erb, R. Libanori, N. Rothfuchs, A.R. Studart, Composites reinforced in three dimensions by using low magnetic fields, Science 335 (2012) 199–204.

[32] A. Sarkar, M.A. Somashekara, M.P. Paranthaman, M. Kramer, C. Haase, I.C. Nlebedim, Functionalizing magnet additive manufacturing with in-situ magnetic field source, Additive Manufacturing 34 (2020) 101289. https://doi.org/10.1016/j.addma.2020.101289.

[33] M.C. Mapley, G. Tansley, J.P. Pauls, S.D. Gregory, A. Busch, Selective laser sintering of bonded anisotropic permanent magnets using an in situ alignment fixture, RPJ ahead-of-print (2021). https://doi.org/10.1108/RPJ-09-2020-0220.